# Characterization of R-134a Superheated Droplet Detector for Neutron Detection

**Prasanna Kumar Mondal** [*,†], **Rupa Sarkar and Barun Kumar Chatterjee**

Department of Physics, Bose Institute, 93/1 A. P. C. Road, Kolkata 700009, India.
e-mail: prasanna_ind_82@yahoo.com (P.K. Mondal)
  sarkar_rupa2003@yahoo.com (R. Sarkar)
  barun_k_chatterjee@yahoo.com  (B.K. Chatterjee)

[*] Corresponding author:
  Tel.: +913323031185; fax: +913323506790.
  e-mail: prasanna_ind_82@yahoo.com (P.K. Mondal)

**Abstract**
R-134a ($C_2H_2F_4$) is a low cost, easily available and chlorine free refrigerant, which in its superheated state can be used as an efficient neutron detector. Due to its high solubility in water the R-134a based superheated droplet detectors (SDD) are usually very unstable unless the detector is fabricated using a suitable additive, which stabilizes the detector. The SDD is known to have superheated droplets distributed in a short-lived and in a relatively longer-lived metastable states. We have studied the detector response to neutrons using a $^{241}$AmBe neutron source and obtained the temperature variation of the nucleation parameters and the interstate kinetics of these droplets using a two-state model.

**1. Introduction**
The radiation detectors based on the emulsion of superheated liquid droplets in viscoelastic gel or in soft polymer matrix are being used for neutron detection, neutron dosimetry and neutron spectrometry for over three decades (Apfel et al., 1982; d'Errico et al., 2001, 2002; Mukherjee et al., 2007). Compared to the other conventional neutron detectors the superiority of these detectors has already been well established. In superheated droplet detector (SDD), the active liquid is dispersed in the form of micron-size droplets in a viscoelastic gel medium. It is well known that the superheated state is a metastable state of the liquid, where a small perturbation like, the thermal fluctuations, energy deposition by energetic radiation etc., could trigger the formation of a stable vapour phase. The SDD is used in almost all branches of radiation physics, including health physics, medical physics, space physics, nuclear physics and high energy physics (Apfel et al., 1989; Harper et al., 1995; Ing et al., 1999; Guo et al., 1999). Superheated liquid based detectors are also used in dark matter search experiments (Behnke et al., 2011; Archambault et al., 2012; Felizardo et al., 2012), since by choosing the operating temperature and pressure it can be made completely insensitive to majority of the backgrounds associated with these experiments. COUPP (Behnke et al., 2011), PICASSO (Archambault et al., 2012) and SIMPLE (Felizardo et al., 2012) are the three

[†] Present address: Department of Chemical, Biological & Macromolecular Sciences,
      S.N. Bose National Centre for Basic Sciences,
      Block JD, Sector III, Salt Lake, Kolkata 700064, India.
      e-mail: prasanna@bose.res.in.

groups working on the dark matter search experiments using superheated liquid based detectors.

For the preparation of SDD different low boiling point liquids are used, such that at the operating temperature it can be used for the detection of ionizing radiations. The R-12 ($CCl_2F_2$; b.p. - 29.8$^o$C) based SDDs are well studied and are widely used in neutron detection (d'Errico et al., 2002; Mondal et al., 2013). However due to its chlorine content it has an ozone depleting potential (Solomon, 1999) and hence it is banned in many countries. R-134a ($C_2H_2F_4$) has a boiling point of -26.6$^o$C and in SDD it can be used as a possible alternative of R-12 (Harper et al., 1995; Das et al., 2010). We have studied the temperature variation of the nucleation parameters of R-134a SDD prepared by a modified SDD fabrication technique (Mondal and Chatterjee, 2013) that produces a highly stable detector which can be used even years after its fabrication. Recent study has indicated that in SDD there exist two groups of droplets one with a much shorter lifetime than the other (Sarkar et al., 2008). The decay of these droplets is modeled using a *two-state* decay scheme (Mondal and Chatterjee, 2009). We have used this model for fitting the neutron irradiation data of R-134a SDD. The experiments are performed in the temperature range of 20 to 40$^o$C, where for neutron irradiation a 3 Ci $^{241}$AmBe neutron source is used. The nucleation rate data is fitted with the *two-state* model, which gives the nucleation parameters, nucleation efficiency and interstate transition kinetics of the R-134a droplets.

**2. Theory**

The superheated state is a metastable state of the liquid, where the liquid phase is maintained either at a temperature higher than its boiling point at a given pressure or at a pressure lower than its saturation vapour pressure at a given temperature (Avedisian, 1985). In superheated liquid there exists a dynamic population of microbubbles, which grow to a maximum size and then collapse back. If a microbubble reaches a size larger than a certain critical size ($r_c$), bubble nucleation occurs, and then the vapour bubble grows spontaneously vaporising the superheated liquid. Here an energy barrier ($W$) (Roy et al., 1987), due to the interplay between the surface and volume forces, governs the stability of the microbubbles. When a microbubble has sufficient energy to overcome this barrier it causes spontaneous homogeneous bubble nucleation. The bubble nucleation may also be triggered by the energetic radiation which deposits energy in the liquid and causes radiation induced nucleation. For bubble nucleation to occur, the energy deposition has to be greater than the threshold energy ($W$) needed to form a critical size ($r_c$) microbubble. For neutrons the recoil ion, produced due to the neutron-nucleus elastic scattering, deposits energy along its path and triggers the bubble nucleation. The frequency of spontaneous nucleation is usually quite low compared to the radiation induced nucleation, which enables one to use SDD as a radiation detector.

When a SDD is exposed to energetic radiation the superheated droplets vaporise independent of each other. The vaporisation of superheated droplet is accompanied by the emission of an acoustic pulse and a change in volume, both of which can be detected electronically (Apfel and Roy, 1983; Mondal and Chatterjee, 2008). In a radiation field the superheated droplets expected to decay monotonically. However, when a SDD is irradiated multiple times with a radiation-off period in between two irradiations, the observed nucleation rate at the beginning of the irradiation found to be much higher than the nucleation rate at the end of the previous irradiation (Sarkar et al., 2008). This discrepancy in the nucleation rate data indicates that in SDD the droplets are in two metastable states, the *normal metastable state* and the *second metastable state*. The droplets continuously move from one state to the other and at low nucleation rate the droplet population reaches equilibrium. A droplet in the *normal metastable state* has a lifetime which is much longer than its lifetime in the *second metastable state*.

When irradiated with neutrons the droplets in the *second metastable state* decay much faster than the others in the *normal metastable state*, giving a sharp fall in the nucleation rate, after which the short-lived droplet population decreases considerably and the long-lived droplets mainly contribute in the decay data (Mondal and Chatterjee, 2009). During the radiation-off period the short lived droplets repopulate from the *normal metastable state* resulting in an increase in the nucleation rate at later irradiations. It was observed that $d$, the transition rate from second metastable state to normal metastable state, is larger than the transition rate from normal metastable state to second metastable state $c$ (Mondal and Chatterjee, 2013).

During neutron irradiation the superheated droplets in SDD decay due to spontaneous and induced nucleations. The nucleation frequency of the droplets in normal and second metastable states can be expressed as (Mondal and Chatterjee, 2009),

$$b = b^{spont} + b^{induced} = k_o v + k_1 v \psi \qquad (1)$$

and

$$a = a^{spont} + a^{induced} = q_o v + q_1 v \psi. \qquad (2)$$

Here $v$ is the droplet volume, $\psi$ is the neutron flux, $k_o$, $q_o$ are the spontaneous nucleation rate per unit volume for the normal and second metastable states respectively and $k_1$, $q_1$ are the neutron induced nucleation frequency per unit volume per unit flux for the normal and second metastable states respectively. The first and second terms on the right hand side of Eqs. (1) and (2) account for the spontaneous and neutron induced nucleations respectively.

The nucleation parameters $k_o$, $q_o$, $k_1$ and $q_1$, and the interstate transition rates $c$ and $d$ can be obtained by fitting the multi-exposure nucleation rate data with the *two-state* decay model (Mondal and Chatterjee, 2009, 2013).

### 2.1. Nucleation efficiency

The nucleation efficiency $\eta_n$ of a superheated liquid is defined as the probability of bubble nucleation for each scattering event of the neutron (Sarkar et al., 2006). Since different ions have different LETs (linear energy transfer), the probability of triggering a nucleation is also different for different ions. Here, due to the difficulty in identifying which ion has triggered the nucleation it is not possible to calculate the nucleation efficiencies for all the ions separately. Thus we have taken the average nucleation efficiency for all the ions $\eta_n$, which can be expressed as (Sarkar et al., 2006),

$$\eta_n = \frac{k_1}{\frac{\rho_{liq} N_A}{M} \sum_i n_i \sigma_i} \qquad (3)$$

where $\rho_{liq}$ is the liquid density, $M$ is the molecular weight, $N_A$ is the Avogadro number, $n_i$ is the atomicity of the i$^{th}$ nuclear species of the molecule having the neutron-nuclei elastic scattering cross-section $\sigma_i$. Similarly, for the second metastable state the nucleation efficiency $\eta_n^*$ can be written as

$$\eta_n^* = \frac{q_1}{\frac{\rho N_A}{M} \sum_i n_i \sigma_i}. \qquad (4)$$

By finding $k_1$ and $q_1$ the nucleation efficiencies of the liquid $\eta_n$ and $\eta_n^*$ can be obtained using Eqs. (3) and (4).

## 3. Detector fabrication

The R-134a SDD can be prepared by a simple emulsification process (Roy et al., 1998) using R-134a as the active liquid. During the emulsification the liquid breaks into small droplets which remain suspended in a viscoelastic gel medium. The viscoelastic gel is prepared by mixing glycerol and commercial ultrasound gel in a proportion such that it can hold the droplets in suspension. We have observed that the R-134a based SDD is usually very unstable due to the high water solubility of R-134a (0.15 wt% at 1 bar and 25°C), which results in a diffusion of the liquid into the gel gradually vanishing the droplets. In order to overcome this problem we have prepared the SDD by a modified SDD fabrication technique (Mondal and Chatterjee, 2013). In this technique before the emulsification we have added Tween 80 surfactant in the gel, which enhances the stability of the R-134a droplets in the viscoelastic gel. A polydisperse emulsion of superheated droplets (about 8500 drops/ml) is obtained by this method. Here the droplet size distribution (Fig. 1) is measured using a technique reported by Mondal et. al., 2010. In Fig. 1(b) $f(v)$ represents the normalized droplet volume distribution of the emulsion, which is used in fitting the nucleation rate data of R-134a SDD. It is observed that in R-134a SDD the droplet size varies in the range of about 15 to 90 μm.

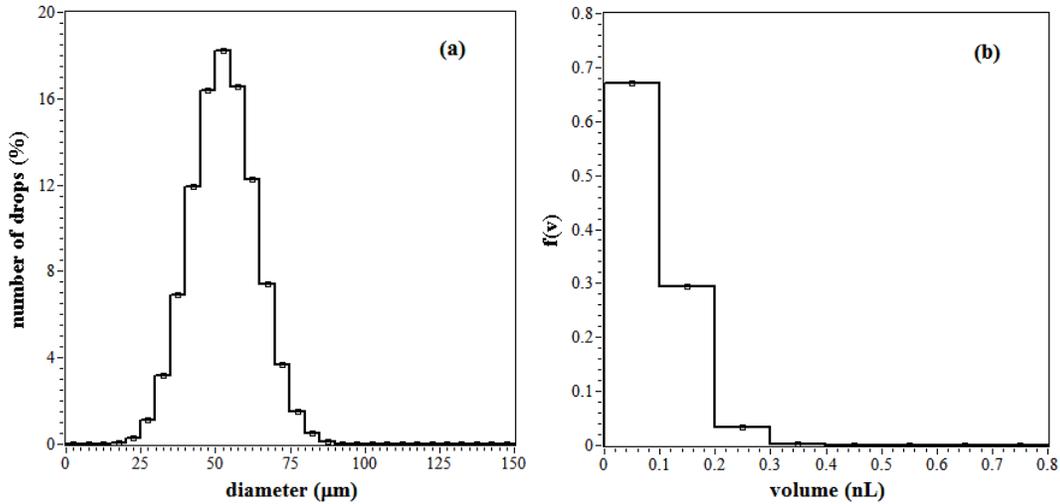

**Fig. 1** The droplet size distribution in R-134a SDD.

We have also done a quantitative comparison of the stability of surfactant free and surfactant added R-134a SDDs by measuring the change in droplet population with the detector ageing for the two cases. For this study we have prepared two batches of SDDs, one is surfactant free and other one is surfactant added. In both the cases same amount of viscoelastic gel and R-134a liquid were used for emulsification and 10 vials of SDDs were prepared in each cases. The vials were stored at about 4°C and were used at different ageing time for counting the number of droplets present in the vial. The total number of droplets present in each vial were measured experimentally by vaporising all the droplets present in the vials. In a typical experiment the vial was wrapped with a heating coil, using which the detector temperature was increased in small steps up to a temperature such that all the droplets are vaporised. The acoustic pulse generated during vaporisation of a superheated droplet was converted into an electric pulse by a $BaTiO_3$ piezoelectric transducer. These electric pulses were converted into TTL pulses with the help of a pulse shaping device (Sarkar et al., 2006) and were counted using a data acquisition card (Advantech USB 4711) operating in a LabView platform. The total number of counts recorded in such an experiment gives the total number of droplets initially present in the vial. The variation in droplet population as a function of ageing of the surfactant free and surfactant added SDDs is shown in Fig. 2. Since the volume of the detector was different in different vials the droplet population is normalized with the detector volume.

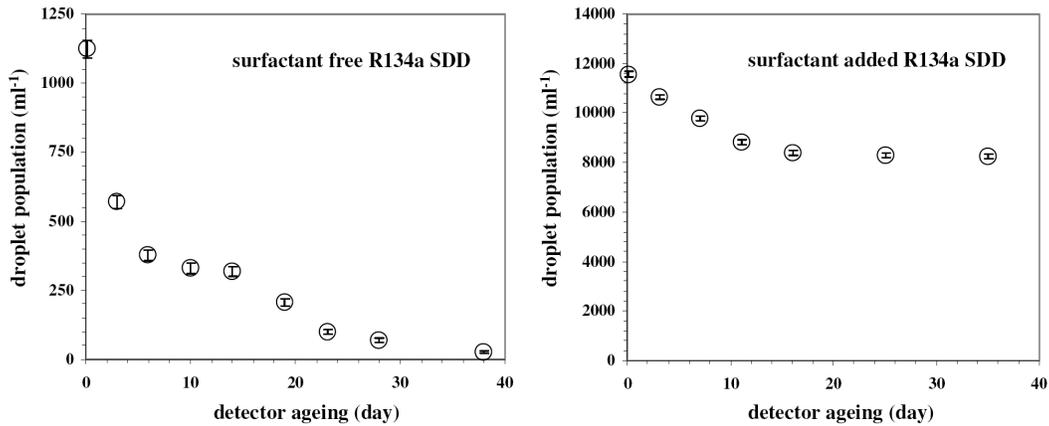

**Fig. 2** Variation of the normalized droplet population in surfactant free and surfactant added R-134a SDD.

## 4. Experimental method

For the characterization of R-134a SDD the experiments were performed at different temperatures using a 3 Ci $^{241}$AmBe neutron source. The schematic diagram of the experimental setup is shown in Fig. 3. In these experiments about 8 ml SDD was taken in a glass vial and was placed on the top of a $BaTiO_3$ piezoelectric transducer. The vial was wrapped with a heating coil, which was connected to a variac using which the detector temperature was controlled. The detector temperature was increased in small steps up to a desired temperature at which the SDD was irradiated with neutrons. The neutron produces recoil ion which, while passing through the active liquid, deposits energy along its path and induces the bubble nucleation when sufficient energy is deposited within a certain critical length (Roy et al., 1987). As discussed earlier the vaporisation of the superheated droplet is associated with the generation of an acoustic pulse, which can be detected by the piezoelectric transducer. With the help of a pulse shaping device (Sarkar et al., 2006) the electric pulse from the piezoelectric transducer is converted into TTL pulse. Using this device we have obtained the nucleation rate data, i.e. the number of droplets vaporised ($N(t)$) during a preset dwell-time at time $t$. The nucleation rate data was acquired as a function of time using a MCS (multichannel scaler) programmed in a LabView platform.

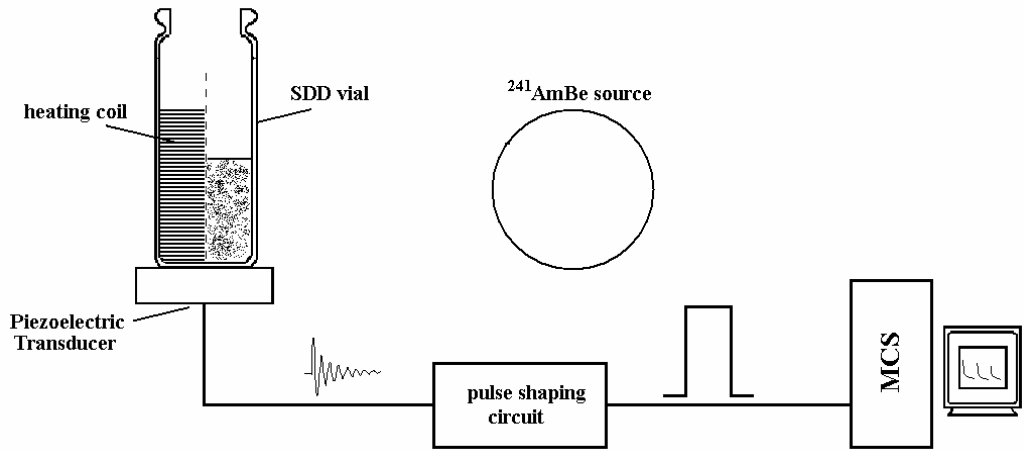

**Fig. 3** Schematic diagram of the experimental setup used for obtaining the nucleation rate data of R-134a SDD using a $^{241}$AmBe neutron source.

In order to obtain the nucleation parameters of two metastable states a multi-exposure nucleation rate data is required (Mondal and Chatterjee, 2009). The two-state fitting of multi-exposure nucleation rate data removes the degeneracy in the obtained nucleation parameters. In our experiments, at a constant temperature and at ambient pressure, initially the R-134a SDD was irradiated with neutrons for a few minutes. The radiation was then turned off for a period of time (by removing the neutron source) and then turned back on again by placing the source. Such switching on and off the irradiation was repeated again where the radiation-off periods were varied. A typical multi-exposure experimental data at temperature 35$^{o}$C is shown in Fig. 4. To understand how the spontaneous and induced nucleation rates change with temperature and also for obtaining the temperature variations of the interstate transition rates the experiments were performed in the temperature range of 20 to 40$^{o}$C. In all these experiments the R-134a SDD was irradiated 3 times with neutrons.

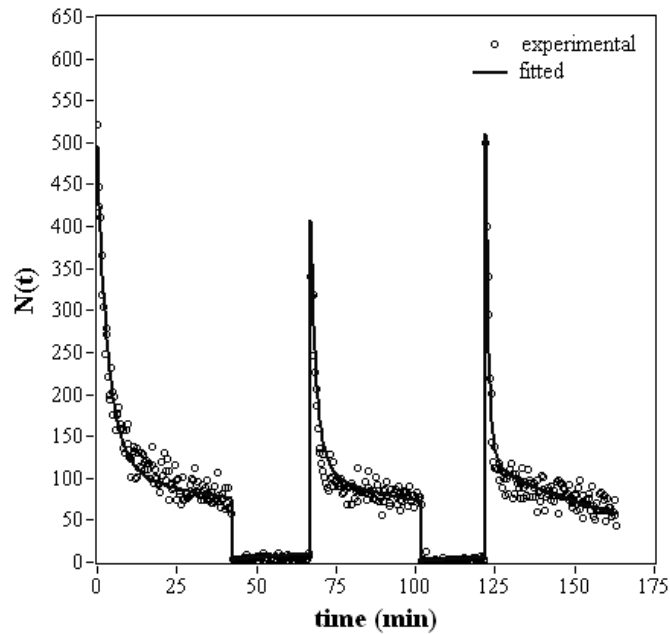

**Fig. 4** A typical experimental and fitted nucleation rate data obtained with R-134a SDD at 35$^{o}$C.

## 5. Results and discussion

The nucleation rate data of R-134a SDD were fitted using the two-state model. The detail of the data fitting method has been reported earlier (Mondal and Chatterjee, 2009). Since the nucleation frequencies $b$ and $a$ (Eqs. 1-2) is dependent on the droplet volume, the droplet volume distribution $f(v)$ plays an important role in the data fitting (Sarkar et al., 2004). Here we have used the measured droplet volume distribution $f(v)$, shown in Fig. 1. By fitting the data one can obtain the parameters $k_o$, $q_o$, $k_1$, $q_1$, $c$ and $d$. Using these parameters one can also obtain other quantities, like the nucleation efficiencies, equilibration time ( $1/(c+d)$ ) and droplet population in different metastable states. A typical experimental and fitted data for R-134a SDD is shown in Fig. 4. The temperature variation of the parameters obtained by fitting the nucleation rate data are shown in Figs. 5-10, which show some characteristic features as discussed bellow.

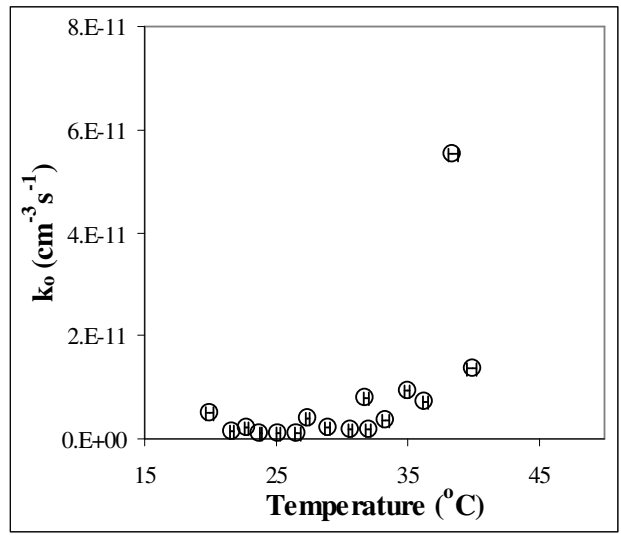

**Fig. 5** The temperature variation of the spontaneous nucleation frequency ($k_o$) for normal metastable state.

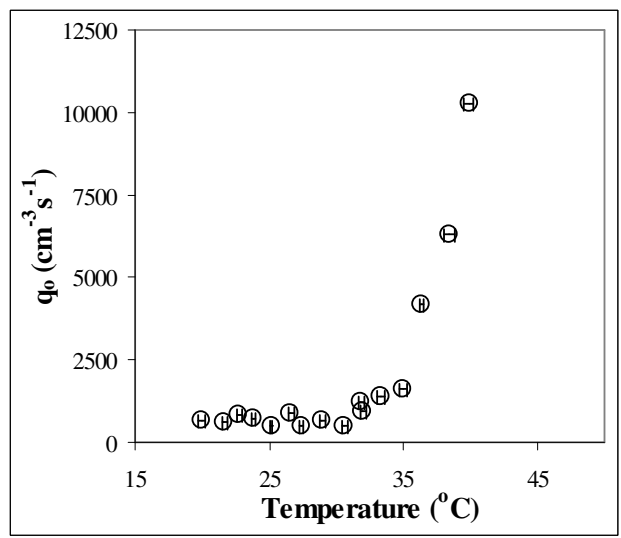

**Fig. 6** The temperature variation of the spontaneous nucleation frequency ($q_o$) for second metastable state.

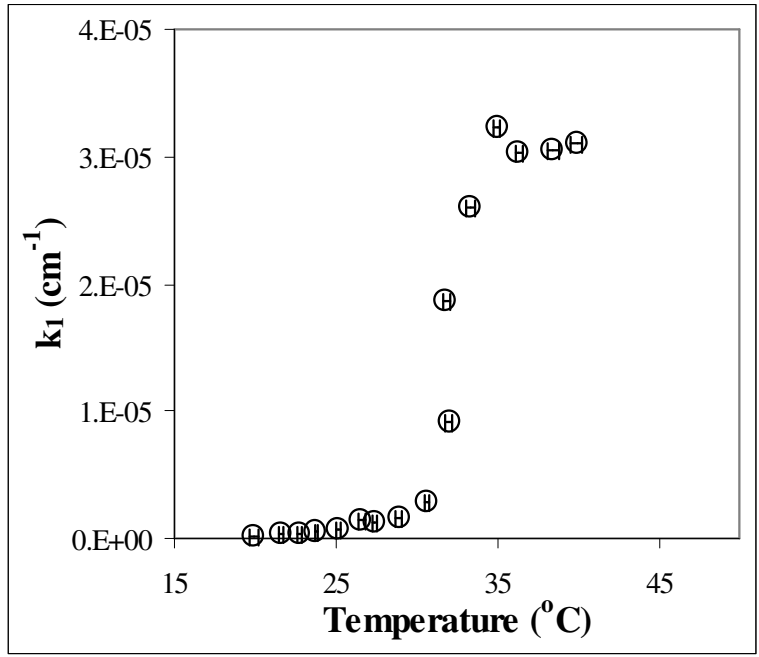

**Fig. 7** The temperature variation of the neutron induced nucleation frequency ($k_1$) for normal metastable state.

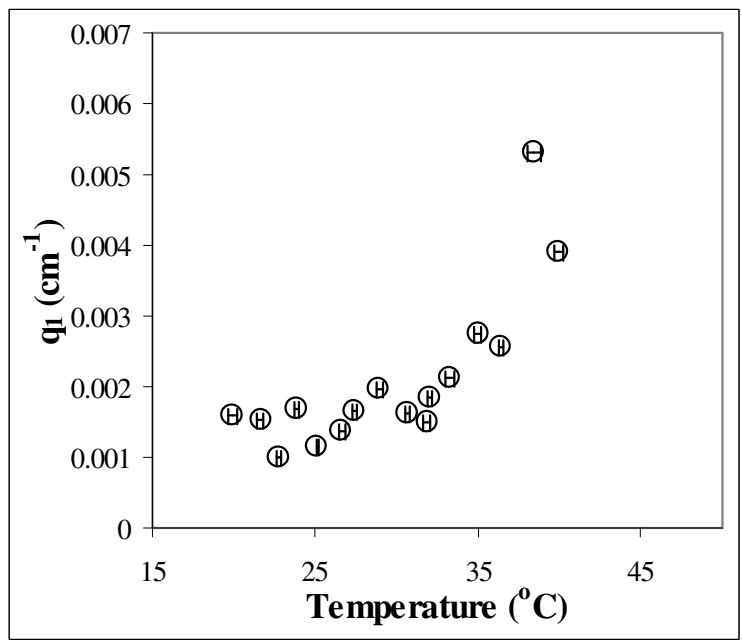

**Fig. 8** The temperature variation of the neutron induced nucleation frequency ($q_1$) for second metastable state.

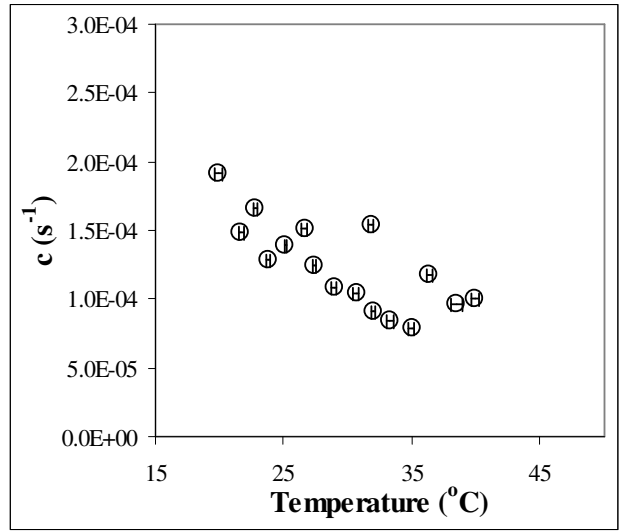

**Fig. 9** The temperature variation of the transition rate ($c$) from normal to second metastable state.

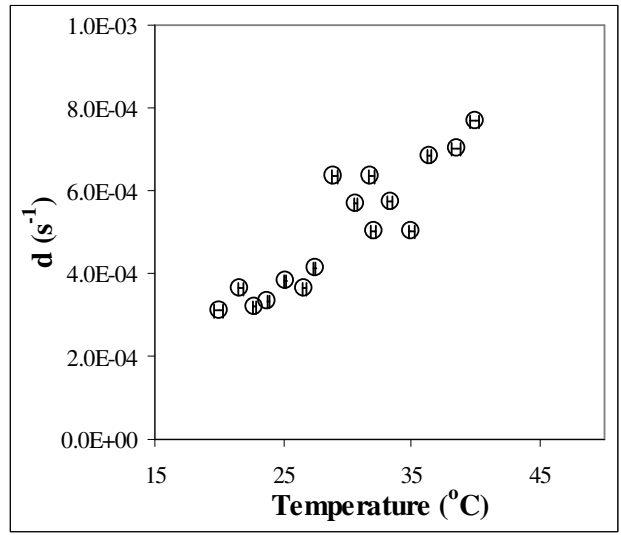

**Fig. 10** The temperature variation of the transition rate ($d$) from second to normal metastable state.

The temperature variations of $k_o$ and $q_o$, the spontaneous nucleation frequency per unit volume of the active liquid for normal and second metastable states respectively, are shown in Fig. 5 and Fig. 6. It is observed that $k_o$ and $q_o$ increase with increase in temperature. This happens because with increase in temperature the threshold energy for nucleation ($W$) decreases while the number of microbubbles per unit volume of the active liquid increases making the superheated liquid more and more unstable. For this reason with increase in detector temperature the probability of bubble nucleation increases, resulting in an increase in $k_o$ and $q_o$.

The neutron induced nucleation frequencies of the two metastable states, $k_1$ and $q_1$, also increase with increase in temperature, as shown in Figs. 7-8. Here $k_1$ and $q_1$ are the neutron induced nucleation frequencies per unit volume of the active liquid for normal and second metastable states respectively. In case of neutron induced events, in addition to the threshold energy and microbubble density, the neutron-nucleus interaction cross-section and the LET affect the nucleation rate. As the threshold energy for nucleation decreases with a rise in detector temperature, more and more ions contribute to the nucleation, resulting in an increase in $k_1$ and $q_1$.

The temperature variation of the transition rates $c$ and $d$ are shown in Figs. 9-10. It is observed that with an increase in temperature, $c$ decreases, while $d$ increases. This indicates that, as the temperature increases the probability of transition from normal to second metastable state decreases while the probability of transition from second to normal metastable state increases. For this reason the droplet population in second metastable state decreases with an increase in temperature, as shown in Fig. 11. At the equilibrium the fraction of droplets present in second metastable state is $P_S = c/(c+d)$ (Mondal and Chatterjee et al., 2009, 2013). Using $c$ and $d$, the temperature dependence of $P_S$ is obtained (Fig. 11).

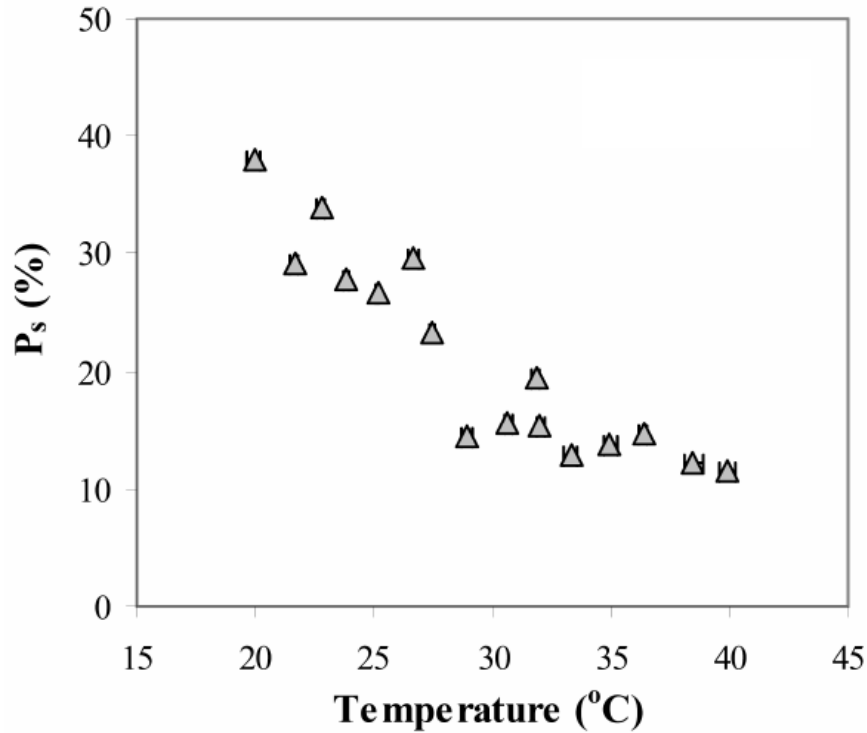

**Fig. 11** Temperature variation of the short-lived droplet population in R-134a detector.

The equilibration time of the system can be estimated using the transition rates $c$ and $d$. In absence of any nucleation events after a time $\tau = 1/(c+d)$, known as the equilibration time, the system will reach equilibrium and then there will be no substantial change in the droplet populations of different metastable states. The temperature variation of $\tau$ is shown in Fig. 12, which shows that with increase in temperature the droplet populations equilibrate among themselves in shorter times.

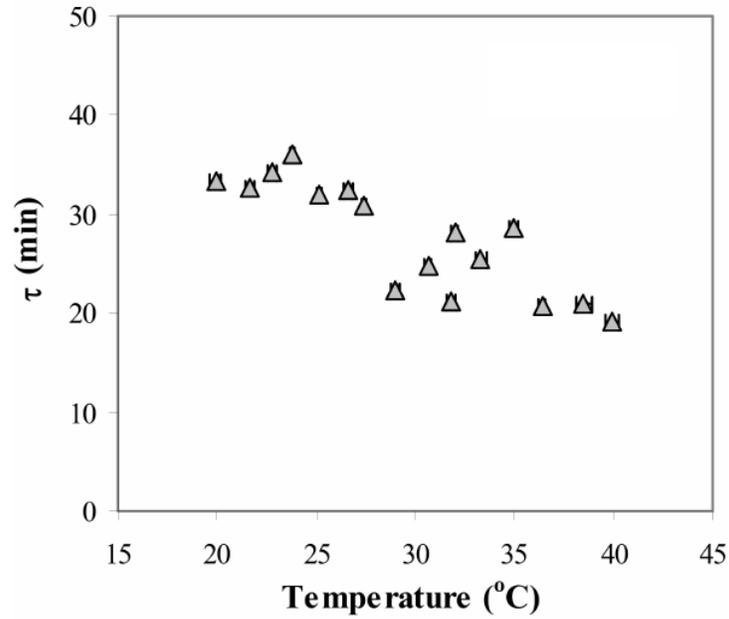

**Fig. 12** Temperature variation of the equilibration time $\tau$.

As discussed earlier, using $k_1$ and $q_1$ values in Eqs. 3-4 one can obtain the nucleation efficiencies $\eta_n$ and $\eta_n^*$ of the liquid in normal and second metastable state respectively. The temperature variations of $\eta_n$ and $\eta_n^*$ for R-134a are shown in Figs. 13-14, which indicate that the probability of bubble nucleation for a droplet in second metastable state is always higher than that in normal metastable state.

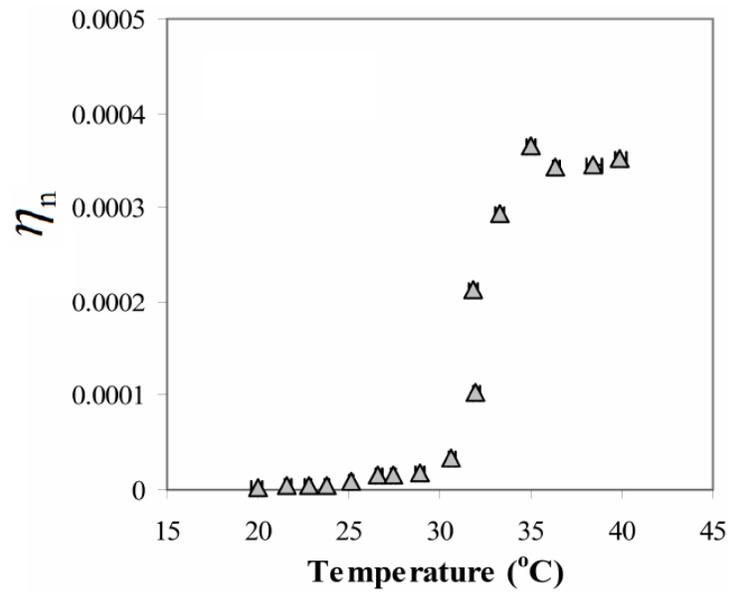

**Fig. 13** The temperature variation of the nucleation efficiency ($\eta_n$) for normal metastable state.

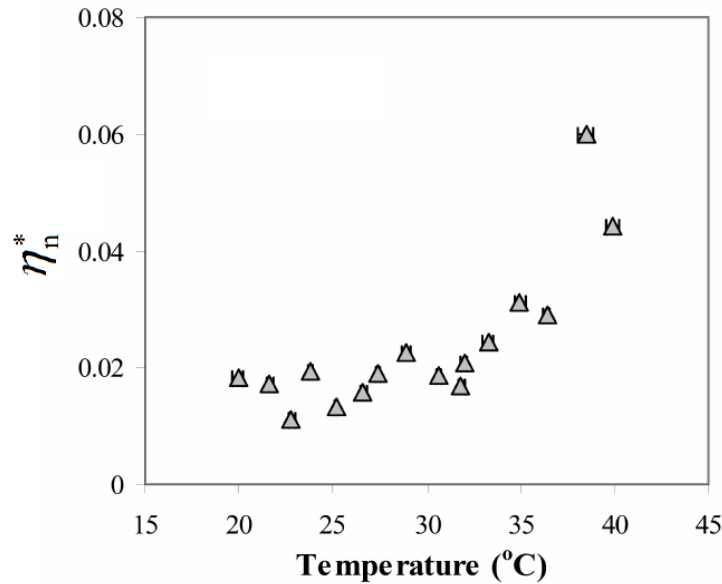

**Fig. 14** The temperature variation of the nucleation efficiency ($\eta_n^*$) for second metastable state.

## 5. Conclusion

We have fabricated a stable R-134a SDD which is neutron sensitive at room temperature. The response of this SDD is studied using a $^{241}$AmBe neutron source. Experimental result shows a strong presence of second metastable state in R-134a SDD. The temperature variation of the nucleation parameters and interstate transition kinetics of the superheated droplets are studied in the temperature range of 20 to 40°C. The large nucleation efficiencies ($\eta_n$ and $\eta_n^*$) per unit volume of R-134a make it a good neutron detector.